\documentclass[prl,english,reprint,longbibliography,superscriptaddress, breaklinks=true, showkeys, showpacs=false, nofootinbib]{revtex4-2}

\usepackage[T1]{fontenc}
\usepackage[utf8]{inputenc}
\usepackage{color}
\usepackage{babel}
\usepackage{amsmath}
\usepackage{amsthm}
\usepackage{amssymb}
\usepackage{subfigure}
\usepackage{graphicx}
\usepackage{physics} 
\usepackage{dsfont}
\usepackage{hyperref}
\usepackage{float}
\usepackage{tabularray}
\usepackage{tabularx}
\usepackage{array}
\usepackage{subcaption}
\usepackage{ragged2e}

\usepackage[dvipsnames]{xcolor}
\usepackage{mathrsfs}
\usepackage{comment}

\begin{document}

\title{Geometry of restricted information: the case of quantum thermodynamics}

\author{Tiago Pernambuco}
\email{tiago4iece@gmail.com}
\affiliation{Theoretical and Experimental Physics Department, Federal University of Rio Grande do Norte, 59078-970, Natal, Brazil}

\author{Lucas C. C\'eleri\href{https://orcid.org/0000-0001-5120-8176}{\includegraphics[scale=0.05]{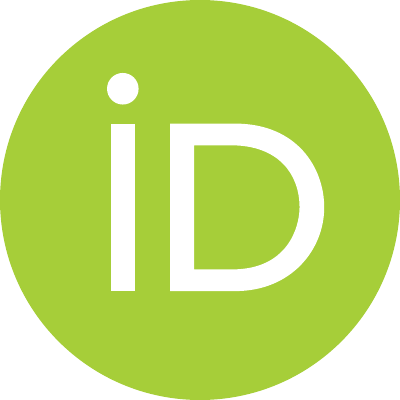}}}
\email{lucas@qpequi.com}
\affiliation{QPequi Group, Institute of Physics, Federal University of Goi\'as, Goi\^ania, Goi\'as, 74.690-900, Brazil}
\affiliation{Instituto de Física de São Carlos, Universidade de São Paulo, CP 369, 13560-970, São Carlos, SP, Brasil}

\begin{abstract}
We formulate a geometric framework in which physical laws emerge from restricted access to microscopic information. Measurement constraints are modeled as a gauge symmetry acting on density operators, inducing a gauge-reduced space of physically distinguishable states. In the case of quantum thermodynamics, this construction leads to a gauge-invariant formulation in which the invariant entropy admits a stochastic description and satisfies a general detailed fluctuation theorem. From this result, we derive an integrated fluctuation theorem and a Clausius-like inequality that unifies the first and second laws in terms of invariant work and coherent heat. Entropy production is identified with the relative entropy between forward and backward probability measures on the gauge-reduced space of thermodynamic trajectories, revealing irreversibility as a geometric consequence of limited observability. The third law emerges as a singular zero-temperature limit in which thermodynamic orbits collapse and entropy production vanishes. Since the framework applies to arbitrary information constraints, it encompasses energy-based thermodynamics as a particular case of more general measurement scenarios.
\end{abstract}

\maketitle

%%%%%%%%%%%%%%%%%%%%%%%%%%%%%%%%%%%%%%%%%%%%%%%%%%%%%%%%%%%%%%%%%%%%%%%%%%%%%%%%
%%%%%%%%%%%%%%%%%%%%%%%%%%%%%%%%%%%%%%%%%%%%%%%%%%%%%%%%%%%%%%%%%%%%%%%%%%%%%%%%
%%%%%%%%%%%%%%%%%%%%%%%%%%%%%%%%%%%%%%%%%%%%%%%%%%%%%%%%%%%%%%%%%%%%%%%%%%%%%%%%
\section{Introduction} 

Physical laws are ultimately formulated relative to the information accessible to an observer, and thermodynamics provides a paradigmatic example of how such restrictions give rise to irreversibility. Thermodynamics is fundamentally based on a coarse-grained description of physical systems. Although microscopic dynamics are governed by reversible laws, macroscopic observations inevitably discard detailed information, leading to the emergence of thermodynamic variables and irreversibility. In classical physics, this transition is often justified by statistical arguments, such as the central limit theorem, which suppresses relative fluctuations in large systems~\cite{Callen1991}. In contrast, many standard formulations of quantum thermodynamics assume near-complete control over the state of the system and define thermodynamic quantities using full microscopic information~\cite{Goold2016,Parrondo2015,Sagawa2013,Deffner2019,Strasberg2022} (see also~\cite{Campbell2026} and references therein). This raises a fundamental question: how can irreversibility emerge in quantum systems when only limited information is operationally accessible?

Recent works have addressed this question by modeling restricted access to quantum degrees of freedom as a gauge symmetry acting on the space of density operators~\cite{ThermoGauge1,ThermoGauge2,Pernambuco2025}. In this framework, the density operator is viewed as a carrier of information, part of which is operationally redundant given the measurement capabilities of the agent. For instance, if the agent can measure only energy, the information associated with coherences or basis choices within degenerate eigenspaces becomes physically irrelevant. This redundancy is removed by the action of a thermodynamic gauge group, producing gauge-invariant quantities that faithfully represent the accessible thermodynamic information. Gauge reduction thus provides a fundamental and geometric implementation of coarse-graining in quantum thermodynamics~\cite{ThermoGauge1,ThermoGauge2,Pernambuco2025}.

In the present article, we demonstrate that the invariant entropy introduced in~\cite{ThermoGauge2} admits a fully consistent stochastic trajectory formulation and satisfies a general detailed fluctuation theorem. This establishes the compatibility of the gauge-invariant framework with fluctuation-theorems that underlies modern nonequilibrium thermodynamics. From this result, we derive an integrated fluctuation relation and a generalized Clausius inequality that unifies the first and second laws within the gauge-reduced setting. Entropy production is identified with the relative entropy between forward and backward probability measures on the gauge-reduced path space, revealing irreversibility as a geometric consequence of restricted observability. Because the construction applies to arbitrary sets of accessible observables, standard energy-based quantum thermodynamics emerges as a particular instance of a more general theory, and the resulting fluctuation relations extend beyond existing formulations~\cite{Esposito2009,Jarzynski2011,Seifert2012,Campisi2011,Jarzynski1997,Jarzynski1997b,Crooks1998,Crooks1999,Crooks2000}. In this way, the present work elevates the geometric formulation of restricted information to a complete nonequilibrium theory grounded in symmetry principles.
 
%%%%%%%%%%%%%%%%%%%%%%%%%%%%%%%%%%%%%%%%%%%%%%%%%%%%%%%%%%%%%%%%%%%%%%%%%%%%%%%
%%%%%%%%%%%%%%%%%%%%%%%%%%%%%%%%%%%%%%%%%%%%%%%%%%%%%%%%%%%%%%%%%%%%%%%%%%%%%%%%
%%%%%%%%%%%%%%%%%%%%%%%%%%%%%%%%%%%%%%%%%%%%%%%%%%%%%%%%%%%%%%%%%%%%%%%%%%%%%%%%
\section{Gauge theory of quantum thermodynamics}

We focus on the case where the agent has access only to energy measurements, leading to a geometric formulation of quantum thermodynamics. Different choices of accessible observables define different thermodynamic theories through changes in the underlying gauge group, while geometric construction and methods remain unchanged~\cite{ThermoGauge1,ThermoGauge2,Pernambuco2025}.

The theory of quantum thermodynamics proposed in Refs.~\cite{ThermoGauge1,ThermoGauge2} received a rigorous geometric formulation in terms of fiber bundles in Ref.~\cite{Pernambuco2025}. We employ this formulation to unravel the geometric structure behind stochastic quantum thermodynamics.

Consider a quantum system described by a time-dependent density operator $\rho_t$ and a Hamiltonian $H_t$. When accessible measurements are restricted to energy measurements, different density operators may become thermodynamically indistinguishable. This redundancy defines an emerging thermodynamic gauge group $\mathrm{G}_T$, whose action leaves all physically accessible quantities invariant. In the case of energy measurements, $\mathrm{G}_T$ is given by $\mathrm{G}_T(t) \simeq \mathrm{U}(n^1_t)\times \mathrm{U}(n^2_t)\times \cdots \times \mathrm{U}(n^k_t)$, where $n^i_t$ are the degeneracies of the instantaneous Hamiltonian eigenvalues and $\sum_i n^i_t=d$, with $d$ the Hilbert-space dimension. Physical thermodynamic functionals are required to be invariant under the action $\rho_t \mapsto V_t \rho_t V_t^\dagger$, with $V_t \in \mathrm{G}_T(t)$.

To formalize this structure, two distinct but related geometric objects are needed. First, a trivial principal $U(d)$-bundle over time $\xi = (\mathbb{R}\times U(d), \pi, U(d), \mathbb{R})$ is introduced, where the base space $\mathbb{R}$ is time and the fiber is the full unitary group $U(d)$ acting on the Hilbert space of the system. A connection in this bundle is defined by the (right-invariant) Maurer--Cartan form $A_t = \dot{u}_t u_t^\dagger$, where $u_t$ diagonalizes $H_t$. This connection induces the covariant derivative $\nabla_t(\cdot) = \partial_t(\cdot) + [A_t,\,\cdot]$, which plays the role of a gauge-covariant time derivative.

Within this framework, time-dependent Hermitian operators are interpreted as sections of an associated vector bundle with fiber given by the space of Hermitian matrices, transforming under the adjoint representation of $\textrm{U}(d)$. In particular, the density operator $\rho_t$ is treated as a matter field. Gauge-invariant definitions of work and heat naturally follow as~~\cite{Pernambuco2025}
\begin{align}
W_{\mathrm{inv}}[\rho_t] &= \int_0^\tau \dd t\, \mathrm{Tr}\!\left(\rho_t \nabla_t H_t\right),\label{work}\\
Q_{\mathrm{inv}}[\rho_t] &= \int_0^\tau \dd t \, \mathrm{Tr}\!\left(H_t \nabla_t \rho_t\right)\label{heat}.
\end{align}

One detail of particular importance to the present work is that, by defining the parallel transport of a density matrix in the associated bundle through the usual equation $\nabla_t\rho_t = 0$, we see that it corresponds to paths without invariant heat exchange. For a closed system, where no heat is exchanged between the system and some environment, $Q_{\mathrm{inv}} = Q_c$, where $Q_c$, called coherent heat, is associated with the generation of coherences and is defined as~\cite{ThermoGauge1,ThermoGauge2}
\begin{equation}
    Q_c[\rho] = \int_0^\tau dt\Tr (\rho \dot u_t h_t u_t^\dagger + \rho u_th_t \dot u_t^\dagger),
\end{equation}
where $h_t = u_t H_t u^{\dagger}_t$. Thus, following the arguments presented in~\cite{ThermoGauge1,ThermoGauge2}, parallel transport in closed systems corresponds to paths where no information about the system is lost in the form of effective heat, thus defining the adiabatic transformations under restricted information.

A second geometric structure is associated with the group $\mathrm{G}_T(t)$. Since $\mathrm{G}_T$ depends explicitly on time, it does not define a single principal bundle in $\mathbb{R}$. Instead, it gives rise to a family of trivial principal bundles over individual time instants (or periods), each equipped with its own Maurer-Cartan connection. This structure encodes the coarse-graining induced by the measurement constraints and formalizes the emergent nature of the thermodynamic gauge symmetry.

Gauge-invariant entropy is defined by group averaging (quantum twirling) over $\mathrm{G}_T$,
\begin{equation}
S_{\mathrm{G}_T}[\rho_t] = S_{\mathrm{vN}}(\rho_t^{E}),
\label{entropy}
\end{equation}
where $\rho_t^{E}$ is block-diagonal in the energy eigenspaces and is maximally mixed within each degenerate subspace (see Supplemental Material, Section I, and Refs.~\cite{ThermoGauge1,ThermoGauge2,Pernambuco2025} for further details and explicit expressions). More generally, any unitary-invariant functional $F$ gives rise to a gauge-invariant thermodynamic quantity through $F_{\mathrm{inv}}[\rho_t]=F[\rho_t^{E}]$.

The bundle structure provides a unifying geometric language for quantum thermodynamics, placing it on the same conceptual footing as modern gauge theories and opening the door to topological and geometric methods in the study of thermodynamic processes.

%%%%%%%%%%%%%%%%%%%%%%%%%%%%%%%%%%%%%%%%%%%%%%%%%%%%%%%%%%%%%%%%%%%%%
%%%%%%%%%%%%%%%%%%%%%%%%%%%%%%%%%%%%%%%%%%%%%%%%%%%%%%%%%%%%%%%%%%%%%
%%%%%%%%%%%%%%%%%%%%%%%%%%%%%%%%%%%%%%%%%%%%%%%%%%%%%%%%%%%%%%%%%%%%%
\section{The fluctuation theorem} 

The stochastic nature of quantum thermodynamics allows for a detailed analysis through fluctuation theorems~\cite{Jarzynski1997,Crooks1999}. Within the framework of gauge theory, we must construct these theorems using trajectories that respect the thermodynamic symmetry defined by $\mathrm{G}_T$, since they are the only ones accessible to the agent. 

Consider a thermodynamic process that occurs in some interval of time $t\in [0,\tau]$. Projective energy measurements at $t=0$ and $t=\tau$ define gauge-invariant trajectories~\cite{Note1}. A measurement of the energy eigenvalue $\epsilon^k_t$ projects the system onto a subspace associated with the projector $\Pi_{n_t^k}$. The fundamental assumption of the theory is that we lack access to microscopic degrees of freedom within degenerate subspaces. Consequently, conditioned on the outcome $\epsilon^k_t$ the state of the system must be the gauge-twirled state (the maximally mixed state of that subspace) $\rho^{k}  = \Pi_{n_t^k}/n_t^k$~\cite{ThermoGauge2}.

Irreversibility arises when it is possible for us to tell the time-forward process from the backward process. The stochastic entropy production, which quantifies this distinguishability, is defined as the natural logarithm of the ratio between the probability associated to these processes. Let the forward process ($F$) begin in an arbitrary gauge-invariant state $\rho_F = \sum_k p^k_F \rho^k$ (which includes equilibrium states) where $p^k_F$ is the probability of finding the system with energy $\epsilon^k_0$. Following the measurement of $\epsilon^k_0$, the system unitarily evolves under $U_\tau$ (defined by $H_t$). The probability of measuring the eigenvalue $\epsilon^l_{\tau}$ (with degeneracy $n_\tau^l$) at time $\tau$ is given by the transition probability $p(l|k) = \Tr(\Pi_{n_\tau^l} U_\tau \rho^k U_\tau^\dagger)$, from which follows the joint probability of the forward process $p_F(k,l) = p^k_F p(l|k)$.

For the reverse process ($R$), we define a gauge-invariant reference state $\rho_R = \sum_l p^l_R \rho^l$. For example, the thermal state associated with the Hamiltonian $H_{\tau}$. The time-reverse evolution is governed by $U_\tau^\dagger$~\cite{Note2}. The joint probability of observing $\epsilon^l_{\tau}$ and then $\epsilon^k_0$ is given by $p_R(l,k) = p^l_R p(k|l)$, with the reverse conditional probability given by $p(k|l) = \Tr(\Pi_{n_\tau^k} U_{\tau}^{\dagger} \rho^l U_{\tau})$.

From Eq.~\eqref{entropy}, we define the stochastic gauge-invariant entropy associated with a measurement outcome $k$ at time $t$ as (see Supplemental Material, Section I)
\begin{equation}
    s(k) = -\ln\left(\frac{p^k}{n_t^k}\right).
\end{equation}
This definition captures two distinct contributions to the uncertainty: the classical probability distribution ($-\ln p^k$) and the intrinsic quantum uncertainty due to degeneracy ($\ln n_t^k$), which leads to the Holevo asymmetry discussed in Ref.~\cite{ThermoGauge2}.

The stochastic entropy production $\sigma_{\mathrm{inv}}$ for a single trajectory is the difference between the final and initial stochastic entropies $\sigma_{\mathrm{inv}} = s_R(l) - s_F(k)$, which directly leads to the detailed fluctuation relation for agents who are restricted to energy measurements
\begin{equation}
    \sigma_{\mathrm{inv}} = \ln\left(\frac{p^k_F}{p^l_R}\right) + \ln\left(\frac{n_\tau^l}{n_0^k}\right).
    \label{eq:detailed}
\end{equation}
The standard Crooks fluctuation theorem~\cite{Crooks1999} is a direct consequence of this result.

Geometrically, an energy measurement does not select a quantum state, but rather an entire orbit of states within a degenerate eigenspace under the action of $\mathrm{G}_T$. The states belonging to the same orbit are thermodynamically indistinguishable since they are related by gauge transformations. Consequently, $p^k$ represents the probability assigned to an orbit, rather than to a point in the Hilbert space, with the factor $1/n_t^k$ corresponding to the normalized measure induced by group averaging. This is the coarse-graining implemented by restricted measurements.

Equation~\eqref{eq:detailed} shows that irreversibility has two distinct geometric origins. The term involving the ratio of degeneracies quantifies the entropy production arising from the loss of information associated with changes in degeneracy along the process. This contribution vanishes for trivial thermodynamic subgroups, but represents an intrinsically geometric source of irreversibility whenever degeneracies are present.

The second contribution persists even in the absence of degeneracy. In geometric terms, the choice of initial and reference states $\rho_F$ and $\rho_R$ defines distinct sections of the associated bundle~\cite{Pernambuco2025}. Under the same dynamics, these sections induce different probability measures in the reduced (gauge-invariant) space of trajectories. Entropy production then quantifies the mismatch between these forward- and backward-path measures, providing a geometric origin of irreversibility that does not rely on degeneracy.

It is important to observe that the detailed fluctuation theorem holds for any gauge-invariant reference state, reflecting the fact that entropy production is defined as a relative entropy between forward and backward path measures. However, its standard thermodynamic interpretation in terms of free energy differences requires choosing the reference state as the thermal (Gibbs) state associated with the final Hamiltonian. With this choice, degeneracy factors cancel, entropy production reduces to the familiar work–free energy relation, and the Clausius inequality follows directly. Other choices of reference state correspond to more general nonequilibrium cases and do not lead to standard thermodynamic potentials.

In summary, degeneracy-related irreversibility arises from the loss of information within subspaces that are inaccessible to the agent, leading to an additional entropic contribution. Non-degenerate irreversibility, on the other hand, originates from the asymmetry between forward and backward ensembles induced by state preparation and driving. Both effects admit a unified geometric interpretation in terms of inequivalent probability measures generated by transport in the thermodynamic associated bundle.

From Eq.~\eqref{eq:detailed} we directly obtain the integral fluctuation theorem 
\begin{equation}
    \langle e^{-\sigma_{\mathrm{inv}}} \rangle = 1,
    \label{IFT}
\end{equation}
which, by Jensen's inequality, results in the non-negativity of the entropy production
\begin{equation}
    \langle \sigma_{\mathrm{inv}} \rangle = S_{\mathcal{G}_T}(\rho_R) - S_{\mathcal{G}_T}(\rho_F) \ge 0.
    \label{eq:positivity}
\end{equation}

Equation~\eqref{IFT} expresses the integrated fluctuation theorem as a purely geometric identity that relates two probability measures defined on the same reduced space of thermodynamic trajectories. After quotienting the Hilbert space by the thermodynamic gauge group, each trajectory corresponds to a curve in the space of energy-equivalent states. The forward and backward processes induce two absolutely continuous measures (with respect to each other) on this gauge-invariant path space, associated with the same projected dynamics but different probability densities determined by state preparation and time reversal. Equation~\eqref{IFT} states that these measures are related by a change in probability density, with the exponential of entropy production playing the role of the Radon–Nikodym derivative. As such, the identity reflects the normalization of the backward measure when pulled back along forward trajectories and is independent of degeneracy, driving, or distance from equilibrium. It is therefore a kinematic statement about the geometric structure of the reduced path space rather than a manifestation of irreversibility.

Equation~\eqref{eq:positivity} follows by averaging the logarithm of the Radon–Nikodym derivative and identifies the mean entropy production with the relative entropy between the forward and backward measures on the gauge-invariant path space. Consequently, Eq.~\eqref{eq:positivity} expresses irreversibility as the impossibility of making the forward and backward probability measures coincide after gauge reduction, except in the special case of global detailed balance. Physically, this means that even when microscopic dynamics is reversible, the combination of coarse-graining, gauge symmetry, and state preparation induces an intrinsic asymmetry between forward and backward ensembles. Geometrically, entropy production quantifies the mismatch between probability measures induced by inequivalent sections and measures transported along the same connection.

The detailed fluctuation theorem implies the generalized Clausius inequality (Supplemental Material, Sec. III)
\begin{equation}
    W_\mathrm{inv} \ge \Delta \mathcal F^{eq} + \frac{1}{\beta} \Delta S_{\mathcal{G}_T} + Q_{{c}}[\rho_\tau].
    \label{eq:claussius}
\end{equation}
This inequality sets a lower bound on the invariant work. We should remember that the usual second law states that $W_u \ge \Delta \mathcal F^{eq}$, where, in our formalism, $W_u = W_{\mathrm{inv}} - Q_c$ is the average work as usually defined in quantum thermodynamics~\cite{Alicki1979,ThermoGauge2}. Due to the restriction imposed on the possible measurements, Eq.~\ref{eq:claussius} states a stronger lower bound, taking into account the increase in (invariant) entropy and the loss of energy in the form of coherent heat. This is a direct consequence of the coarse-grained imposed on the information space by the thermodynamic group.   

Geometrically, Eq.~\eqref{eq:claussius} shows that, once gauge reduction is taken into account, the minimum work principle acquires an intrinsically geometric correction. Work is required not only to change the equilibrium free energy on the base space but also to accommodate changes in the entropy induced by degeneracy variations and by the exchange of coherent heat. In this sense, Eq.~\eqref{eq:claussius} unifies the first and second laws: irreversibility is encoded in the nontrivial geometry of the thermodynamic group, while energetic costs are associated with motion on the gauge-reduced space of thermodynamic states, thus reflecting changes in the gauge-invariant quantities (such as energies) and with changes in the invariant volume along gauge orbits.

This result is derived in Section II of the Supplemental Material directly from the gauge-invariant quantities, without making reference to the fluctuation theorem, thus providing a clear proof of the internal consistency of the theory. We also relate this inequality with a quantum coherence measure, making the connection between irreversibility and coherence generation even more explicit.

%%%%%%%%%%%%%%%%%%%%%%%%%%%%%%%%%%%%%%%%%%%%%%%%%%%%%%%%%%%%%%%%%%%%%%%
%%%%%%%%%%%%%%%%%%%%%%%%%%%%%%%%%%%%%%%%%%%%%%%%%%%%%%%%%%%%%%%%%%%%%%%
%%%%%%%%%%%%%%%%%%%%%%%%%%%%%%%%%%%%%%%%%%%%%%%%%%%%%%%%%%%%%%%%%%%%%%%
\section{The third law} 

Let us consider a quantum system described by the Gibbs state $\rho_\beta$ defined by the inverse temperature $\beta$ and a time dependent Hamiltonian $H_t$. We first write the thermal state in the energy eigenbasis in terms of the energy projectors $\Pi_{n^k_t}$ and explicitly separate the contributions from the ground state manifold. Then, by taking the limit $T \rightarrow 0$, we obtain $\rho_{\beta\rightarrow \infty} = \Pi_{n_{t}^{0}}/n_{t}^{0}$, which is the maximally mixed state in the ground state manifold. The gauge-invariant entropy results in $S_{\mathcal \mathrm{G}_T}[\rho]_{T=0} = \ln n_{t}^{0}$. Thus, when the ground state is non-degenerate, the gauge-invariant entropy at $T=0$ vanishes. When it is degenerate, it has a constant value $\ln n_{t}^{0}$.

Within this framework, the third law acquires a clear geometric interpretation as a singular limit within the thermodynamic state space. As the temperature approaches zero, the thermodynamic gauge group collapses, and the accessible states reduce to the ground-state orbit, causing the gauge-reduced state space to become trivial or singular. In this limit, the gauge-invariant path space collapses, eliminating distinguishable thermodynamic trajectories and forcing both forward and backward measures to concentrate on a single point. As a result, entropy production necessarily vanishes, not due to enhanced reversibility of the dynamics but because the geometric structure required to support irreversibility ceases to exist. The unattainability principle then corresponds to the impossibility of reaching this singular region of the state space through any finite, physically allowed transport, providing a geometric formulation of the third law consistent with both thermodynamic and dynamical considerations.

%%%%%%%%%%%%%%%%%%%%%%%%%%%%%%%%%%%%%%%%%%%%%%%%%%%%%%%%%%%%%%%%%%%%%%%
%%%%%%%%%%%%%%%%%%%%%%%%%%%%%%%%%%%%%%%%%%%%%%%%%%%%%%%%%%%%%%%%%%%%%%%
%%%%%%%%%%%%%%%%%%%%%%%%%%%%%%%%%%%%%%%%%%%%%%%%%%%%%%%%%%%%%%%%%%%%%%%
\section{Conclusions} 

The present work advances the gauge-theoretic formulation of quantum thermodynamics in several directions. We introduce a stochastic notion of gauge-invariant entropy defined at the level of individual trajectories, which naturally separates classical uncertainty from contributions arising due to degeneracies. Based on this, we formulate a detailed fluctuation theorem directly on the gauge-reduced space, where trajectories correspond to equivalence classes under the thermodynamic gauge group, thereby explicitly incorporating the effects of restricted information. Within this framework, entropy production acquires a clear interpretation as a relative entropy between forward and backward path measures, providing a geometric and information-theoretic origin for irreversibility. We further derive a generalized Clausius inequality that unifies the first and second laws while accounting for coherent heat and entropy contributions associated with degeneracy. Finally, we show that the third law emerges naturally as a geometric singular limit in which thermodynamic orbits collapse, rendering the reduced state space trivial and forcing entropy production to vanish.

More generally, the framework is not restricted to energy measurements. By selecting a different observable, or a set of observables, one defines a corresponding thermodynamic gauge group and, consequently, a distinct thermodynamic description constructed within the same geometric structure. For instance, in a trapped-ion platform, one may consider energy measurements for the vibrational degrees of freedom while allowing full access to the internal (spin) degrees of freedom. This leads to a modified set of thermodynamic quantities and relations, reflecting the specific measurement constraints, while preserving the underlying geometric construction. In this sense, the present work does not propose a single alternative formulation of quantum thermodynamics, but rather a unified framework from which different thermodynamic theories emerge depending on the accessible observables.

Within this setting, the fluctuation theorems acquire a purely geometric meaning. Forward and backward processes define absolutely continuous probability measures in a space of a reduced trajectory space based on a common gauge, and the integrated fluctuation theorem expresses their normalization under a change in probability density. Entropy production is identified with the relative entropy between these measures, so that irreversibility emerges as the distinguishability of thermodynamic path ensembles rather than as a consequence of microscopic dissipation alone.

This perspective yields a unified geometric formulation of the thermodynamic laws: the zeroth law corresponds to the existence of a consistent gauge symmetry; the first law to the invariant decomposition of energy changes into work and heat; the second law to the positivity of relative entropy on the reduced path space; and the third law to a singular zero-temperature limit in which gauge orbits collapse and entropy production vanishes. Thus, irreversibility appears as a structural consequence of symmetry and limited observability.

Moreover, within the present framework, entropy production is identified with a relative entropy between forward and backward path measures. For small deviations, this admits the expansion of the invariant entropy in terms of the Fisher information, the Hessian matrix of the relative entropy. From Cramér-Rao inequality, one directly obtains a thermodynamic uncertainty relation, exactly as first derived in~\cite{Micadei2013} as far as we know, although not under this name. This shows that thermodynamic uncertainty relations arise naturally as a consequence of the geometric and information-theoretic structure of the reduced path space.

Because the framework applies to arbitrary measurement constraints, it is directly relevant to experimental platforms. In particular, the framework predicts an energetic cost associated with changes in the degeneracy structure of the Hamiltonian, which can arise even in the absence of coherence generation, as well as a separation between coherent heat and other energetic contributions that allows one to distinguish between energy stored in populations and energy associated with inaccessible coherent degrees of freedom. Additionally, the geometric contribution to dissipation, related to the distinguishability between the gauge-reduced state and the corresponding equilibrium state, introduces a correction that persists even when coherence is suppressed. Such effects can be probed in controlled quantum platforms, such as trapped-ion systems~\cite{Bouton2021,Pijn2022,Yan2018,Onishchenko2024}, photonic setups~\cite{Oliveira2020,Ribeiro2020,Zanin2019}, and many-body quantum engines~\cite{Chen2019,Koch2023,Watanabe2020}, where energy measurements can be implemented with high precision and coherent dynamics can be engineered and selectively controlled, enabling direct access to the different contributions predicted by the theory.

More broadly, the gauge-theoretic structure shared with fundamental physics naturally opens directions toward relativistic quantum thermodynamics~\cite{Costa2026,Moreira2024,Basso2025,Basso2023} and related extensions.

%----------------------------------
\vspace{0.2cm}\emph{Acknowledgments}. This research was funded by CNPq through grant 308065/2022-0, the National Institute of Science and Technology for Applied Quantum Computing through CNPq grant 408884/2024-0, FAPEG through grant 202510267001843, and FAPESP through grant 2025/23726-4. LCC also acknowledges the warm hospitality of the Instituto de Física de São Carlos.

%%%%%%%%%%%%%%%%%%%%%%%%%%%%%%%%%%%%%%%%%%%%%%%%%%%%%%%%%%%%%%%%%%%%%%%
%%%%%%%%%%%%%%%%%%%%%%%%%%%%%%%%%%%%%%%%%%%%%%%%%%%%%%%%%%%%%%%%%%%%%%%
%%%%%%%%%%%%%%%%%%%%%%%%%%%%%%%%%%%%%%%%%%%%%%%%%%%%%%%%%%%%%%%%%%%%%%%
\appendix

%%%%%%%%%%%%%%%%%%%%%%%%%%%%%%%%%%%%%%%%%%%%%%%%%%%%%%%%%%%%%%%%%%%%%
%%%%%%%%%%%%%%%%%%%%%%%%%%%%%%%%%%%%%%%%%%%%%%%%%%%%%%%%%%%%%%%%%%%%%
%%%%%%%%%%%%%%%%%%%%%%%%%%%%%%%%%%%%%%%%%%%%%%%%%%%%%%%%%%%%%%%%%%%%%
\section{Gauge theory of the thermodynamic group}
\label{app:gaugethermo}

In this Section, we briefly review the gauge theory approach to quantum thermodynamics developed in Refs.~\cite{ThermoGauge1,ThermoGauge2,Pernambuco2025}. We do not aim to be complete, but only to provide the main ideas behind this geometric formulation, its consequences, and also to provide some explicit expressions that are useful in practical calculations.  

From now on, we consider the agent restricted to energy measurements. The formalism also applies to other observables, although the structure of the thermodynamic group will be different. The central premise of this framework is that if an agent is restricted to energy measurements, a distinction between states that yield the same energy statistics is not possible. The internal energy of a system described by the Hamiltonian $H_t$ and density matrix $\rho_t$ is $U[\rho_t] = \Tr(\rho_t H_t)$, which remains invariant under any unitary transformation $V_t$ that commutes with the Hamiltonian. These transformations form the thermodynamic gauge group $\mathrm{G}_T$, which mathematically encodes the indistinguishability of the microstates of the system given the restricted access of information from the agent~\cite{ThermoGauge1}. For a general Hamiltonian with degeneracies, this group is isomorphic to $\mathrm{G}_T = \mathrm{U}(n_t^1) \times \mathrm{U}(n_t^2) \times ... \times \mathrm{U}(n_t^k)$, where $\mathrm{U}(n_t^i)$ represents the unitary group acting on the $n_t^i$-fold degenerate subspace of the $i$-th energy eigenvalue. Note that we can choose any set of observables to define the group. The formalism remains the same.

To construct a thermodynamic theory consistent with this limited access, the physical quantities must be invariant under $\mathrm{G}_T$~\cite{ThermoGauge1} (the gauge principle). Since an arbitrary state $\rho_t$ may contain hidden information (coherences) that changes under gauge transformations, we define the physically accessible state $\rho^\epsilon_t$ by averaging $\rho_t$ over the gauge group. This process, known as twirling, projects the state in a block-diagonal form~\cite{ThermoGauge2}
\begin{equation}
    \rho^E_t = \bigoplus_{k=1}^{p \leq d}\frac{\Tr[\Pi_{n_t^k}\rho_t\Pi_{n_t^k}]}{n_t^k} \mathbb I_{n_t^k},
    \label{rhoddE}
\end{equation}
where $\Pi_{n_t^k}$ projects onto the $k$-th eigenspace, and $\mathbb I_{n_t^k}$ is the identity on that subspace.

Gauge invariant work $W_{\mathrm{inv}}$ and heat $Q_{\mathrm{inv}}$ are defined in terms of this operator. In reality, these definitions are derived from the geometric structure of the theory, in which they are associated with the covariant derivative, the connection in the main fiber bundle, associated with the gauge potential defined by the thermodynamic group~\cite{Pernambuco2025}.

We define gauge-invariant work $W_{\mathrm{inv}}$ as the variation of energy due to changes in the Hamiltonian parameters, evaluated on this effective state~\cite{ThermoGauge1,ThermoGauge2}:
\begin{equation}
W_{inv}[\rho_t] = \int_0^\tau \dd t \Tr(\rho^E_t \dot{H}_t).
    \label{Winv}
\end{equation}
This definition contrasts with the standard definition of average work found in the literature, denoted here as $W_u$, which is calculated using the full state $\rho_t$~\cite{Alicki1979}
\begin{equation}
    W_u[\rho_t] = \int_0^\tau \dd t \Tr(\rho_t \dot{H}_t).
\end{equation}
The difference between these two quantities reveals that the gauge-invariant work captures only the energy exchange associated with population changes and energy level shifts, excluding contributions from coherence. This is natural since the agent is restricted to only measuring energy and thus does not have access to quantum coherences on this basis. These two quantities are related by the expression $W_{\mathrm{inv}}[\rho_t] = W_u[\rho_t] - Q_c[\rho_t]$~\cite{ThermoGauge1}, where
\begin{equation}
Q_{c}[\rho_t] = \int_0^\tau \dd t \Tr(\dot{\rho}^E(t) H_t),
    \label{Qc}
\end{equation}
is called coherent heat and represents the energy cost associated with the generation of quantum coherences --- degrees of freedom that are dynamically active but thermodynamically inaccessible under the gauge constraint~\cite{ThermoGauge1,ThermoGauge2}. 

Similarly, gauge-invariant heat $Q_{\mathrm{inv}}$ can be written as~\cite{ThermoGauge2}:
\begin{equation}
    Q_{\mathrm{inv}}[\rho_t] = Q_{u}[\rho_t] + Q_{c}[\rho_t],
    \label{Qinv}
\end{equation}
where $Q_u = \int_0^\tau \dd t \Tr(\dot{\rho}_t H_t)$ is the usual definition of heat in quantum thermodynamics~\cite{Alicki1979}.

These expressions for $W_{\mathrm{inv}}$IQ1 and $Q_{\mathrm{inv}}$ are equivalent to those presented in the main text in terms of the covariant derivative. In addition, we can obtain the same quantities directly from the Haar measure of the thermodynamic group. We take the Haar average over the thermodynamic group of the integrated power as the invariant work. This is the usual procedure employed in gauge theories in order to define physical quantities. Observing that the energy is invariant under this average, we define invariant heat as the difference between this energy and the work. This shows the internal consistency of the theory.

It is interesting to observe that the coherent heat defined above emerges from the effective dynamics of the accessible density operator $\rho^{E}_t$. Consider a closed quantum system evolving under the action of a time-dependent Hamiltonian $H_t$. The dynamics of $\rho_{t}^{E}$ is given by~\cite{ThermoGauge2}
\begin{equation}
    \dv{\rho_{t}^{E}}{t} = -i[H_t,\rho_{t}^{E}] + \mathcal{L}(\rho_{t}^{E}),
    \label{app:eq:lindblad}
\end{equation}
with $\mathcal{L}(\rho_{t}^{E}) = \int\dd\mathcal{G}_{T}\left(\dot{V}_t\rho_t V_{t}^{\dagger} + V_t\rho_t \dot{V}_{t}^{\dagger}\right)$ encompassing the irreversibility induced by the coarse-graining implemented by the gauge group. This reflects our lack of ability to fully control the system. If we consider an open quantum system, an additional Lindibladian operator will be added to this equation in the usual way.

Finally, the thermodynamic entropy is defined as the von Neumann entropy of the gauge-invariant state~\cite{ThermoGauge2},
\begin{equation}
    S_{\mathrm \mathrm{G}_T}[\rho_t] = -\Tr(\rho^E_t \ln \rho^E_t).
    \label{SG}
\end{equation}
Due to the structure of $\rho^E_t$, which eliminates off-diagonal elements between energy subspaces and maximally mixes within degenerate subspaces, this entropy can be explicitly decomposed into two contributions:
\begin{equation}
    S_{\mathcal{G}_T}[\rho_t] = S_d[\rho_t] + S_{\Gamma}[f_t],
    \label{EntropyDecomp}
\end{equation}
where $S_d[\rho_t]$ is the diagonal entropy (the Shannon entropy of the populations in the energy eigenbasis) and $S_{\Gamma}[f_t] = -\Tr(f_t \ln |f_t|)$ is the Holevo asymmetry measure associated with the matrix $f_t$~\cite{ThermoGauge2}. Note that inside the logarithm, we take the absolute values of the elements of $f_t$. This matrix is constructed to capture the entropic difference between the diagonal and gauge-averaged states, defined as~\cite{ThermoGauge2}:
\begin{equation}
    f_t \equiv \left(\bigoplus_k -\rho_{kk}^E(t) \Pi_1 \right) \bigoplus \left(\bigoplus_k \frac{\Tr\{\rho_{n_t^k}(t)\}}{n_t^k} \mathbb{I}_{n_t^k}\right).
\end{equation}
Here, the first block contains the diagonal elements of the state in the energy basis, and the second block contains the averaged populations within the degenerate subspaces.

Evaluating these terms explicitly, we recover a form that depends on the probabilities $p^k_t = \Tr(\Pi_{n_t^k}\rho_t)$ of occupying the $k$-th energy level and the instantaneous degeneracies $n_t^k$~\cite{ThermoGauge2}:
\begin{equation}
    S_{\mathrm \mathrm{G}_T}[\rho_t] = - \sum_k p^k_t \ln p^k_t + \sum_k p_k(t) \ln n_t^k.
    \label{app:entropy}
\end{equation}
The first term corresponds to the classical uncertainty over the energy levels, while the second term arises from the hidden structure of the degenerate subspaces.

%%%%%%%%%%%%%%%%%%%%%%%%%%%%%%%%%%%%%%%%%%%%%%%%%%%%%%%%%%%%%%%%%%%%%
%%%%%%%%%%%%%%%%%%%%%%%%%%%%%%%%%%%%%%%%%%%%%%%%%%%%%%%%%%%%%%%%%%%%%
%%%%%%%%%%%%%%%%%%%%%%%%%%%%%%%%%%%%%%%%%%%%%%%%%%%%%%%%%%%%%%%%%%%%%
\section{Generalized second law}
\label{app:secondlaw}

In this Section, we derive a generalized second law of thermodynamics from the gauge theory of the thermodynamic group, as discussed in Section~\ref{app:gaugethermo}. This will take the form of a generalized Clausius inequality. The derivation is based on applying Spohn's results for entropy production in open quantum systems~\cite{Spohn} to the effective Lindbladian dynamics for the gauge-invariant state~\eqref{app:eq:lindblad}. 

The entropy produced in the system is quantified by the relative entropy $S(\rho^E_t||\sigma_t) = \Tr(\rho_t^E \ln \rho_t^E) - \Tr(\rho_t^E \ln \sigma_t)$ between the state of the system and the reference thermal state $\sigma_t = e^{-\beta H_t}/Z_t$, $Z_t = \Tr(e^{-\beta H_t})$ being the partition function at time $t$. Explicitly 
\begin{equation}
    S(\rho^E_t||\sigma_t) = -S_{\mathcal \mathrm{G}_T}[\rho_t] + \beta \Tr(\rho_t^E H) + \ln Z_t,
    \label{SRel}
\end{equation}
where the gauge-invariant entropy $S_{\mathcal \mathrm{G}_T}$ arises from its definition in~\eqref{SG} and $\mathcal F^{eq}_t = - \beta^{-1}\ln Z_t$ is the equilibrium free energy.  By defining the invariant non-equilibrium free energy as $\mathcal F_{inv}[\rho_t] = U[\rho_t] - TS_{\mathcal \mathrm{G}_T}[\rho_t]$, Eq.~\eqref{SRel} becomes
\begin{equation}
   S(\rho ^E_t||\sigma_t) = \beta\left(\mathcal F_{\mathrm{inv}}[\rho_t] - \mathcal F^{eq}_t\right),
\end{equation}
Thus, the relative entropy is effectively a measure of the difference in free energy between the state $\rho_t$ and the thermal equilibrium state $\sigma_t$. The positivity of the relative entropy directly implies
\begin{equation}
    \mathcal F_{\mathrm{inv}}[\rho_t] \geq \mathcal F^{eq}_t,
    \label{Finv}
\end{equation}
i.e., the free energy is at a minimum when the system is at thermal equilibrium. It is important to note that, evidently, following the same reasoning with the non-gauge invariant entropy leads to the same inequality but with the non-gauge invariant free energy.

Differentiating equation \eqref{SRel}, we obtain
\begin{eqnarray}
    \dd S(\rho_t^E||\sigma_t) &=& -\dd(S_{\mathcal \mathrm{G}_T}[\rho_t])  + \beta \Tr(\dot \rho_t^EH_t) \dd t \nonumber \\
    &+& \beta \Tr(\rho_t^E \dot H_t)\dd t + \dd(\ln Z_t).
\end{eqnarray}

In Ref.~\cite{ThermoGauge2}, the second term of the right-hand side of this equation is defined as coherent heat
\begin{equation}
    \delta Q_{c}[\rho_t] = \Tr(H_t \dot{\rho}^E_t)\dd t,
\end{equation}
while the third is the gauge-invariant work
\begin{equation}
    \delta W_{\mathrm{inv}}[\rho_t] = \Tr(\dot H_t {\rho}^E_t)\dd t.
\end{equation}
Substituting this into the entropy differential in Eq.~\eqref{SRel} and noting that $W_u[\rho_t] = W_{\mathrm{inv}}[\rho_t] + Q_c[\rho_t]$ \cite{ThermoGauge2}, we obtain
\begin{equation}
\dd S(\rho_t^E||\sigma_t) =  -\dd S_{\mathcal \mathrm{G}_T}[\rho_t] + \beta \delta W_{u}[\rho_t]  + \dd(\ln Z_t).
\label{Balance}
\end{equation}

Integrating~\eqref{Balance} over the duration of the thermodynamic process from $t=0$ to $t=\tau$, and assuming that the system is initialized in a thermal equilibrium state $\rho_0 = \sigma_0 = e^{-\beta H_0}/Z_0$, we obtain the following
\begin{equation}
    \Delta S(\rho_{\tau}^E || \sigma_{\tau}) = -\Delta S_{\mathcal{G}_T} + \beta W_u + \Delta (\ln Z_\tau),
    \label{integrated_2nd_law}
\end{equation}
where $\Delta S_{\mathcal{G}_T} = S_{\mathcal{G}_T}[\rho_\tau] - S_{\mathcal{G}_T}[\rho_0]$ (and similarly for $\Delta \ln Z_\tau$) and we have identified the integrated work term $\int_0^\tau \delta W_u = W_u$. 

Using the positivity of the relative entropy and remembering that the initial state is an equilibrium state, this last equation can be written as
\begin{equation}
    W_u \ge \Delta \mathcal F^{eq} + \frac{1}{\beta} \Delta S_{\mathcal{G}_T}.
    \label{work_bound_intermediate}
\end{equation}
Using invariant work, we have
\begin{equation}
    W_\mathrm{inv} \ge \Delta \mathcal F^{eq} + \frac{1}{\beta} \Delta S_{\mathcal{G}_T} + Q_{c}[\rho_t].
\end{equation}
This inequality sets a lower bound on the work performed on the system.

We can refine the entropy term to explicitly display the quantum contributions arising from the gauge structure. As shown in~\cite{ThermoGauge2}, the gauge-invariant entropy for a system with a possibly degenerate energy spectrum decomposes as
\begin{equation}
    S_{\mathcal{G}_T}[\rho_t] = S_{vn}[\rho_t] + C_{rel}[\rho_t] + S_{\Gamma}[f_t],
    \label{entropy_decomposition}
\end{equation}
where $S_{vn}$ is the von Neumann entropy, $C_{rel}[\rho_t]$ is the relative entropy of coherence~\cite{Baumgratz2014}, and $S_{\Gamma}[f_t]$ is the Holevo asymmetry measure associated with the degeneracies of the Hamiltonian~\cite{Bartlett2003,Marvian2014,ThermoGauge2}.

For a unitary evolution generated by $H_t$, the von Neumann entropy is a constant of motion, $S_{vn}[\rho_\tau] = S_{vn}[\rho_0]$. Furthermore, since the initial state is thermal (diagonal in energy basis), both coherence and asymmetry vanish at $t=0$, which implies $S_{\mathcal{G}_T}[\rho_0] = S_{vn}[\rho_0]$. Substituting Eq.~\eqref{entropy_decomposition} into Eq.~\eqref{work_bound_intermediate}, we obtain
\begin{equation}
    W_\mathrm{inv} \ge \Delta \mathcal F^{eq} + T \left( C_{rel}[\rho_\tau] + S_{\Gamma}[f_\tau] \right) + Q_{c}[\rho_t].
    \label{generalized_second_law}
\end{equation}

%%%%%%%%%%%%%%%%%%%%%%%%%%%%%%%%%%%%%%%%%%%%%%%%%%%%%%%%%%%%%%%%%%%%%
%%%%%%%%%%%%%%%%%%%%%%%%%%%%%%%%%%%%%%%%%%%%%%%%%%%%%%%%%%%%%%%%%%%%%
%%%%%%%%%%%%%%%%%%%%%%%%%%%%%%%%%%%%%%%%%%%%%%%%%%%%%%%%%%%%%%%%%%%%%
\section{Generalized second law: Fluctuation theorem perspective}
\label{App:Derivation}

In this Section, we provide a detailed derivation of the generalized Clausius inequality presented in the main text. We do this by employing our fluctuation theorem. We proceed by analyzing the statistics of the invariant entropy production, $\sigma_{\mathrm{inv}}$, defined within the context of the gauge-invariant fluctuation theorem.

%%%%%%%%%%%%%%%%%%%%%%%%%%%%%%%%%%%%%%%%%%%%%%%%%%%%%%%%%%%%%%%%%%%%%
\subsection{Stochastic Entropy Production}

We consider a thermodynamic process where a system, initially in a thermal equilibrium state $\rho_0 = e^{-\beta H_0}/Z_0$ is driven by a time-dependent Hamiltonian $H_t$ from $t=0$ to $t=\tau$.

A single gauge-invariant trajectory is defined by measuring the energy eigenvalue $\epsilon^k_0$ (with degeneracy $n_0^k$) at the start and $\epsilon^l_\tau$ (with degeneracy $n_\tau^l$) at the end of the process. The stochastic gauge-invariant entropy associated with a measurement outcome $k$ is defined as
\begin{equation}
    s(k) = -\ln p^k + \ln n^k,
\end{equation}
where $p^k$ is the probability that the system being found in the energy level $\epsilon^{k}$.

The entropy produced by the process is the natural logarithm of the ratio between two probability densities: the forward (in time) and backward (time-reverse) processes. If we cannot tell one from the other, the process is thermodynamically reversible. 

The stochastic entropy production for a single forward ($F$) trajectory is the difference between the stochastic entropy of the reference state in the reverse ($R$) process and the initial state in the forward process 
\begin{equation}
    \sigma_{\mathrm{inv}} = s_R(l) - s_F(k) = \ln\left( \frac{n_\tau^l}{p^l_R} \right) - \ln\left( \frac{n_0^k}{p^k_F} \right).
    \label{sigma_def_app}
\end{equation}
The integral fluctuation theorem established in the main text states $\langle e^{-\sigma_{\mathrm{inv}}} \rangle = 1$. Applying Jensen's inequality ($\langle e^{-x} \rangle \ge e^{-\langle x \rangle}$), we obtain the fundamental bound on the average entropy production:
\begin{equation}
    \langle \sigma_{\mathrm{inv}} \rangle \ge 0.
    \label{second_law_stoch}
\end{equation}
To derive the generalized Clausius inequality, we evaluate this average, $\langle \sigma_{inv} \rangle = \sum_{k,l} p_F(k,l) \sigma_{\mathrm{inv}}$, in two complementary ways.

%%%%%%%%%%%%%%%%%%%%%%%%%%%%%%%%%%%%%%%%%%%%%%%%%%%%%%%%%%%%%%%%%%%%%%%%%%
\subsection{Evaluation through work and free energy}

First, we substitute the explicit form of the thermal populations in Eq.~\eqref{sigma_def_app}. For the initial state $\rho_0 = e^{-\beta H_0}/Z_0$, the probability is $p^k_F = n_0^k e^{-\beta \epsilon^k_0}/Z_0$. Similarly, for the reference state $\rho_R = e^{-\beta H_\tau}/Z_\tau$, the probability is $p^l_R = n_\tau^l e^{-\beta \epsilon_l(\tau)}/Z_\tau$. Substituting these into the expression for $\sigma_{inv}$ results in
\begin{equation}
    \sigma_{\mathrm{inv}} = \ln\left( \frac{n_\tau^l Z_\tau}{n_\tau^l e^{-\beta \epsilon^l_\tau}} \right) - \ln\left( \frac{n_0^k Z_0}{n_0^k e^{-\beta \epsilon^k_0}} \right).
\end{equation}
The degeneracy factors cancel out, leaving
\begin{equation}
    \sigma_{\mathrm{inv}} = \left[ \beta \epsilon^l_\tau + \ln Z_\tau \right] - \left[ \beta \epsilon^k_0 + \ln Z_0 \right].
\end{equation}
Rearranging terms, we identify the stochastic energy difference and the equilibrium free energy differences ($\beta F_{eq} = -\ln Z$)
\begin{equation}
    \sigma_{\mathrm{inv}} = \beta \left[ \epsilon^l_\tau - \epsilon^k_0 \right] - \beta \left[ F_{eq}(\tau) - F_{eq}(0) \right].
\end{equation}
The quantity $w^{l,k} = \epsilon^l_\tau - \epsilon^k_0$ represents the energy change of the closed system for this specific trajectory. Taking the average over the ensemble of forward trajectories yields the average usual work $W_u = \langle w \rangle$:
\begin{equation}
    \langle \sigma_{\mathrm{inv}} \rangle = \beta (W_u - \Delta F_{eq}).
    \label{avg_sigma_work}
\end{equation}

%%%%%%%%%%%%%%%%%%%%%%%%%%%%%%%%%%%%%%%%%%%%%%%%%%%%%%%%
\subsection{Evaluation through invariant entropy}

We directly evaluate the average using the definition of the expectation value
\begin{equation}
    \langle \sigma_{\mathrm{inv}} \rangle = \sum_{k,l} p_F(k,l) \left[ s_R(l) - s_F(k) \right].
\end{equation}
Using the marginal probability definition $\sum_l P_F(k,l) = p_k^F$, the average of the initial term is simply the initial gauge-invariant entropy
\begin{equation}
    \sum_{k,l} p_F(k,l) s_F(k) = \sum_k p^k_F s_F(k) = S_{\mathcal{G}_T}[\rho_0].
\end{equation}

For the final term, we note that the actual final probability distribution of the system is $p^l_\tau = \sum_k p_F(k,l)$. The average then becomes
\begin{equation}
    \sum_{k,l} p_F(k,l) s_R(l) = \sum_l p^l_\tau s_R(l) = \sum_l p^l_\tau \ln\left( \frac{n_\tau^l}{p^l_R} \right).
\end{equation}

We now introduce the gauge-invariant entropy of the actual final state $\rho_\tau$, which is given by $S_{\mathcal{G}_T}[\rho_\tau] = \sum_l p^l_\tau \ln(n_\tau^l / p^l_\tau)$. We then write
\begin{eqnarray}
    \langle s_R \rangle &=& \sum_l p^l_\tau \ln\left( \frac{n_\tau^l}{p^l_R} \right) - S_{\mathcal{G}_T}[\rho_\tau] + S_{\mathcal{G}_T}[\rho_\tau] \nonumber\\
    &=& \sum_l p^l_\tau \left[ \ln\left( \frac{n_\tau^l}{p^l_R} \right) - \ln\left(\frac{n_\tau^l}{p^l_\tau} \right) \right] \nonumber \\
    &+& S_{\mathcal{G}_T}[\rho_\tau].
\end{eqnarray}
Combining the logarithms, the degeneracy factors cancel, and we recognize the relative entropy (Kullback-Leibler divergence) between the final gauge-invariant state $\rho_\tau^E$ and the reference thermal state $\sigma_\tau$:
\begin{equation}
    \langle s_R \rangle = \sum_l p^l_\tau \ln\left( \frac{p^l_\tau}{p^l_R} \right) + S_{\mathcal{G}_T}[\rho_\tau] = S(\rho_\tau^E || \sigma_\tau) + S_{\mathcal{G}_T}[\rho_\tau].
\end{equation}
Combining the initial and final averages, we obtain
\begin{equation}
    \langle \sigma_{\mathrm{inv}} \rangle = \Delta S_{\mathcal{G}_T} + S(\rho_\tau^E || \sigma_\tau).
    \label{avg_sigma_entropy}
\end{equation}

%%%%%%%%%%%%%%%%%%%%%%%%%%%%%%%%%%%%%%%%%%%%%%%%%%%%
\subsection{The Inequality}

We equate the two results for $\langle \sigma_{\mathrm{inv}} \rangle$ derived in Eqs.~\eqref{avg_sigma_work} and~\eqref{avg_sigma_entropy}
\begin{equation}
    \beta (W_u - \Delta F_{eq}) = \Delta S_{\mathcal{G}_T} + S(\rho_\tau^E || \sigma_\tau).
\end{equation}

Using Klein's inequality, which guaranties $S(\rho_\tau^E || \sigma_\tau) \ge 0$, we establish the lower bound
\begin{equation}
    W_u \ge \Delta F_{eq} + T \Delta S_{\mathcal{G}_T}.
\end{equation}

Finally, we use the decomposition of the gauge-invariant entropy given in Eq.~\eqref{entropy_decomposition}. Since the system evolves unitarily, the von Neumann entropy is constant ($\Delta S_{vn} = 0$). Additionally, because the process starts in a thermal state (which is diagonal and gauge-invariant), the initial coherence and asymmetry vanish ($C_{rel}[\rho_0] = 0, S_{\Gamma}[f_0] = 0$). Therefore, the entropy change is purely structural
\begin{equation}
    \Delta S_{\mathcal{G}_T} = C_{rel}[\rho_\tau] + S_{\Gamma}[f_\tau].
\end{equation}
Substituting this back into the work bound yields the Generalized Clausius Inequality
\begin{equation}
    W_u \ge \Delta F_{eq} + T \left( C_{rel}[\rho_\tau] + S_{\Gamma}[f_\tau] \right).
\end{equation}

%%%%%%%%%%%%%%%%%%%%%%%%%%%%%%%%%%%%%%%%%%%%%%%%%%%%%%%%%%%%%%%%%%%%%
%%%%%%%%%%%%%%%%%%%%%%%%%%%%%%%%%%%%%%%%%%%%%%%%%%%%%%%%%%%%%%%%%%%%%
%%%%%%%%%%%%%%%%%%%%%%%%%%%%%%%%%%%%%%%%%%%%%%%%%%%%%%%%%%%%%%%%%%%%%
\section{Geometric refinement of the generalized Clausius inequality}

In Section~\ref{App:Derivation} we derived the generalized Second Law of Thermodynamics~\eqref{generalized_second_law}, which bounds the work carried out on the system by the equilibrium free energy difference and the informational costs associated with coherence and asymmetry. That inequality relies on the non-negativity of the relative entropy $S(\rho_{\tau}^{E}||\sigma_{\tau}) \ge 0$. However, recent developments in non-equilibrium quantum thermodynamics allow us to sharpen this bound by interpreting the relative entropy through a geometric lens.

Deffner and Lutz~\cite{DeffnerLutz} derived a generalized Clausius inequality that holds arbitrarily far from equilibrium. They demonstrated that entropy production is bounded from below by the Bures angle $\mathcal{L}$ between the non-equilibrium state and the corresponding equilibrium state. The Bures angle is a thermodynamic length metric defined as:
\begin{equation}
    \mathcal{L}(\rho_1, \rho_2) = \arccos\left(\sqrt{F(\rho_1, \rho_2)}\right),
    \label{bures_angle}
\end{equation}
where $F(\rho_1, \rho_2) = [\Tr(\sqrt{\sqrt{\rho_1}\rho_2\sqrt{\rho_1}})]^2$ is the quantum fidelity. The bound on the relative entropy is given by \cite{DeffnerLutz}:
\begin{equation}
    S(\rho || \sigma) \ge \frac{8}{\pi^2} \mathcal{L}^2(\rho, \sigma).
    \label{deffner_bound}
\end{equation}
We can apply this result directly to our gauge-invariant framework. Returning to \eqref{integrated_2nd_law}, we identified the term $S(\rho_{\tau}^{E}||\sigma_{\tau})$ as the entropic cost of the discrepancy between the final gauge-invariant state $\rho_{\tau}^{E}$ and the thermal state $\sigma_{\tau}$. By substituting the geometric bound (\ref{deffner_bound}) into our work relation, we obtain a geometrically tightened second law:
\begin{equation}
    W_{u} \ge \Delta\mathcal{F}^{eq} + T(C_{rel}[\rho_{\tau}] + S_{\Gamma}[f_{\tau}]) + \frac{8 T}{\pi^2}\mathcal{L}^2(\rho_{\tau}^{E}, \sigma_{\tau}).
    \label{geometric_second_law}
\end{equation}
The term $T(C_{rel} + S_{\Gamma})$ represents the energetic cost of maintaining quantum features, specifically coherences and population asymmetries within degenerate subspaces. 

In contrast, the new geometric term $\frac{8 T}{\pi^2}\mathcal{L}^2(\rho_{\tau}^{E}, \sigma_{\tau})$ acts as a cost to have a final state that, under the constraints of the thermodynamic group, is distinguishable from a thermal equilibrium state.

Therefore, Eq.~\eqref{geometric_second_law} implies that even if the system is fully decohered and symmetric (where $C_{rel} = S_{\Gamma} = 0$), the work is still bounded by a strictly positive geometric term proportional to the distinguishability of the Haar-averaged density matrix from thermal equilibrium. This reinforces the gauge-theoretic perspective: dissipation arises both from the generation of hidden quantum information (gauge non-invariant features) and from the macroscopic displacement from equilibrium (gauge-invariant distance).

Once again, we point out that the notion of work that appears in~\eqref{geometric_second_law} is just the usual work of quantum thermodynamics and, since all terms in the RHS can be calculated from the Hamiltonian and density matrix of the system, this is, again, simply a generalization of the Clausius inequality, providing an even tighter bound than that of \eqref{generalized_second_law}.

Given the strictly geometric nature of this last contribution, it is natural that we should relate it to the fiber-bundle-based construction of the gauge theory of quantum thermodynamics developed in~\cite{Pernambuco2025}. In that work, the authors establish that time dependent density operators can be seen as sections of an associated vector bundle that arises naturally in the theory. In this context, the Bures angle $\mathcal L(\rho_\tau^E, \sigma_\tau)$ measures the distance between these two sections specifically at the point $t = \tau$ on the base space. Since the Bures angle is a Riemannian metric in the space of density operators~\cite{GeometryOfQuantumStates}, it induces a Riemannian structure in the subspace of the associated bundle that relates to them. However, it is not clear whether it is possible to extend this Riemannian structure to the full associated bundle in any physically meaningful way.

Finally, we discuss the topological nature of the cost of asymmetry $TS_\Gamma$. Since $S_\Gamma$ is a measure of how asymmetric a state is with respect to the thermodynamic group $\mathrm{G}_T$, and the cartesian product structure of the thermodynamic group depends on the distribution of the degeneracies $n^k$ of the Hamiltonian, when a thermodynamic process takes a system through a phase transition that alters $n^k$ , $TS_\Gamma$ can be seen as an energy cost related to changing the topology of the thermodynamic group.

%%%%%%%%%%%%%%%%%%%%%%%%%%%%%%%%%%%%%%%%%%%%%%%%%%%%%%%%%%%%%%%%%%%%%
%%%%%%%%%%%%%%%%%%%%%%%%%%%%%%%%%%%%%%%%%%%%%%%%%%%%%%%%%%%%%%%%%%%%%
%%%%%%%%%%%%%%%%%%%%%%%%%%%%%%%%%%%%%%%%%%%%%%%%%%%%%%%%%%%%%%%%%%%%%
\section{Applications of the generalized Clausius inequality}

To verify the bounds given by Eqs.~\eqref{generalized_second_law} and~\eqref{geometric_second_law}, we performed numerical simulations of two paradigmatic models: the Landau-Zener model and the Curie-Weiss (Collective Ising) Model.

\subsection{Landau-Zener Model}

The paradigmatic Landau-Zener Model describes the dynamics of a single qubit evolving under the Hamiltonian~\cite{LandauZener,PartialLandauZener}
\begin{equation}
    H_t = \frac{\Delta}{2}\sigma_x + \frac{vt}{2}\sigma_z,
    \label{LandauZenerHamiltonian}
\end{equation}
where $\Delta$ is the coupling term, $v$ is the sweep velocity, and $t$ is time. $\sigma_i$ is the Pauli matrix in the $i$ direction.

\begin{figure}[H]
    \centering
    \includegraphics[width=1.0\linewidth]{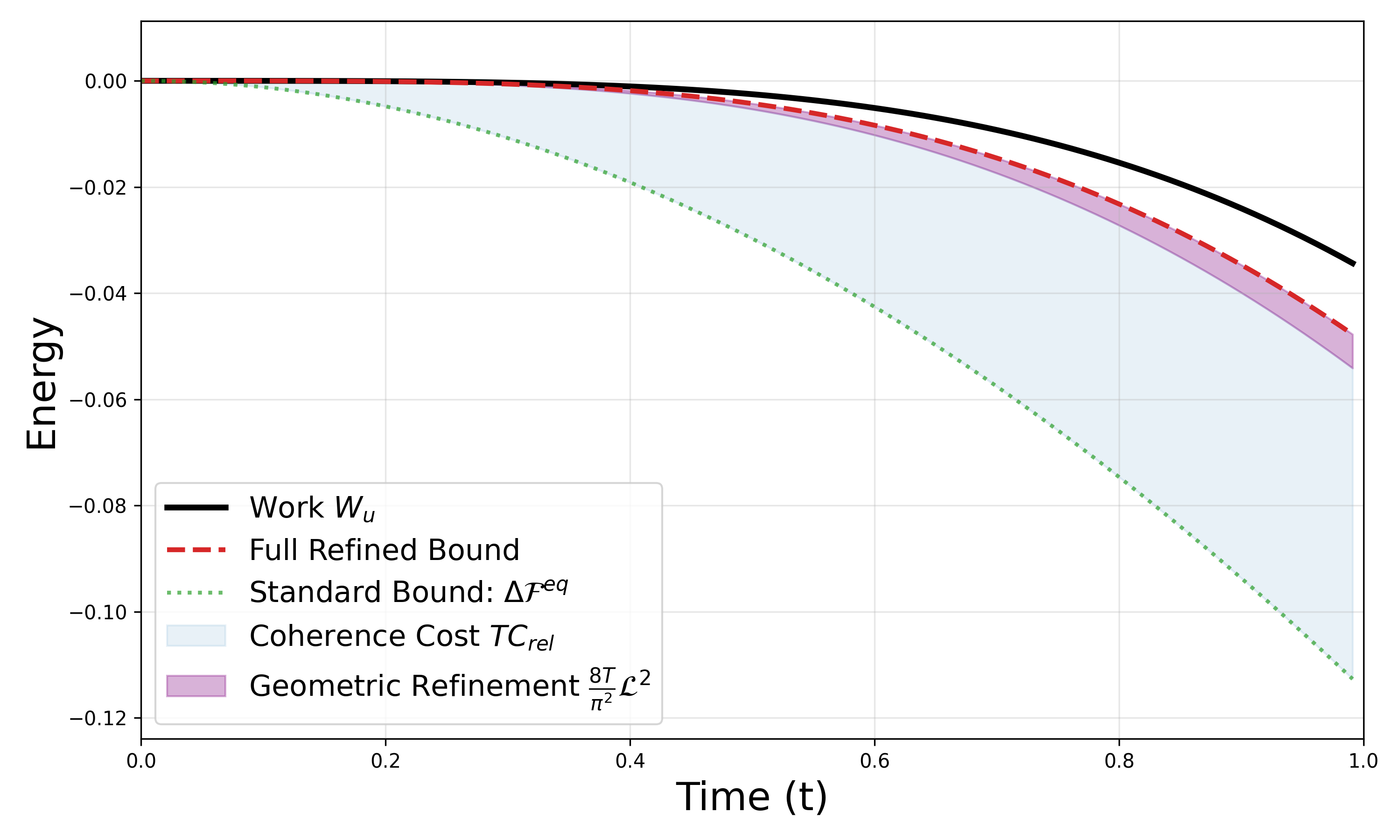}
    \caption{The thermodynamic work $W_u = W_{\mathrm{inv}} - Q_c$ is compared to the bound of eq. \eqref{geometric_second_law}. Particularly, we see that quantum coherence plays a very significant role in tightening the Clausius Inequality for this model.}
    \label{LandauZener}
\end{figure}

For the purpose of verifying Clausius inequalities, we perform the following driving protocol: first, we initialize our qubit in the thermal Gibbs state with $T = 0.5$, $\Delta = 2$, and $v = 1$. We then evolve it in time according to the Hamiltonian~\eqref{LandauZenerHamiltonian} between $t = 0$ and $ t = 1$. We take the Boltzmann constant to unity, thus allowing temperature to be measured in the same units as energy. In addition, we set $\hbar = 1$, which turns the inverse of time into a unit of energy. By choosing an arbitrary unity of time, $t$ is dimensionless, and we also have dimensionless energy, temperature, and velocity. Our results are displayed in Fig.~\ref{LandauZener}.
%%%%%%%%%%%%%%%%%%%%%%%%%%%%%%%%%%%%%%%%%%%%%%%%%%%%%%%%%%%%%%%%%%%%%%%%%%%%%%%%%%
%%%%%%%%%%%%%%%%%%%%%%%%%%%%%%%%%%%%%%%%%%%%%%%%%%%%%%%%%%%%%%%%%%%%%%%%%%%%%%%%%%
It can be clearly seen in Fig.~\ref{LandauZener} that quantum coherence plays a crucial role in the thermodynamics of the system, being responsible for most of the difference between the work performed by the system and the variation in its free energy. That is the origin of irreversibility in the context of gauge thermodynamics. Information is leaking to inaccessible degrees of freedom.

\subsection{Curie-Weiss Model}

\begin{figure}[H]
    \centering
    \includegraphics[width=1.0\linewidth]{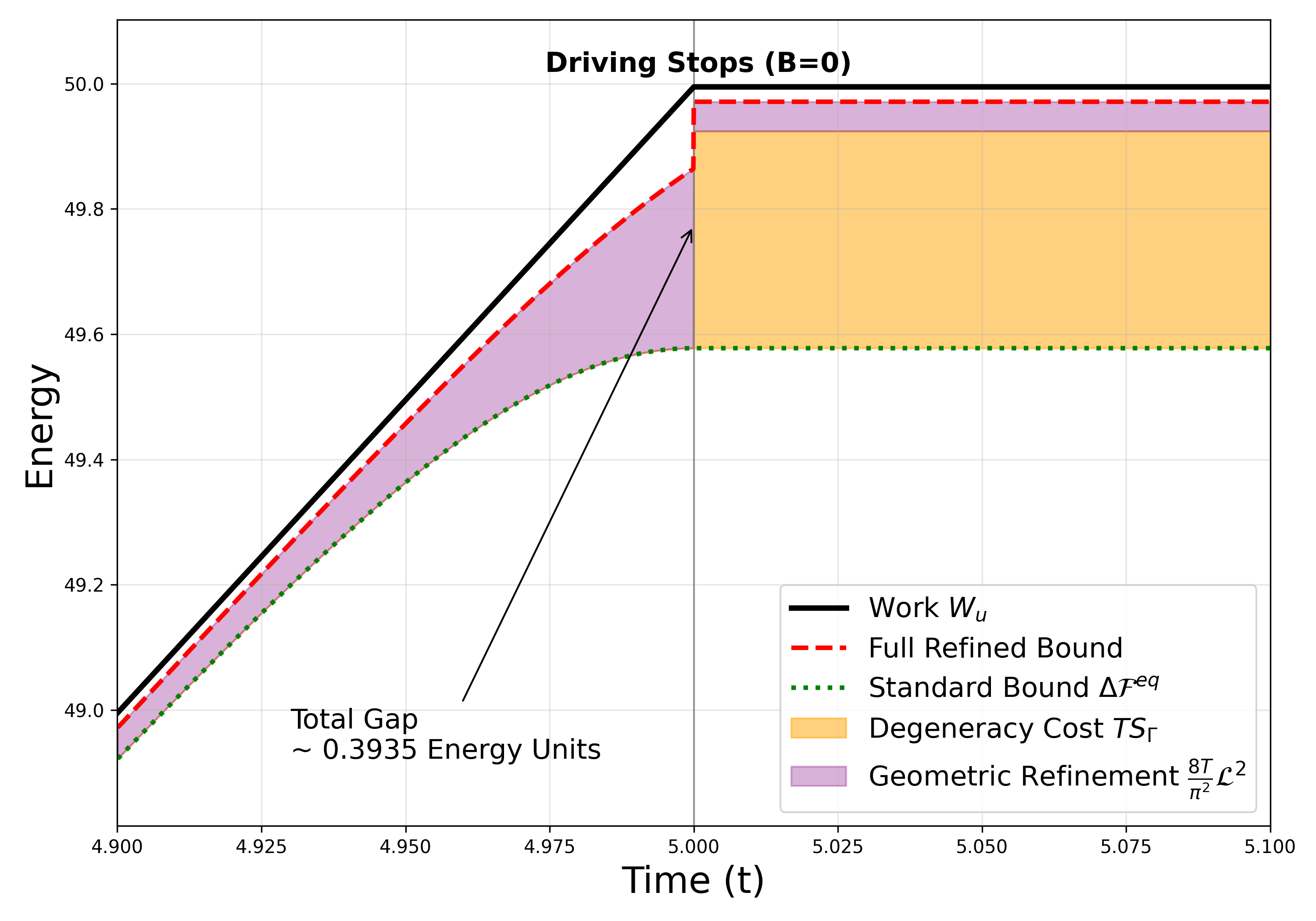}
    \caption{The thermodynamic work $W_u = W_{\mathrm{inv}} - Q_c$ is plotted against the bound of eq. \eqref{geometric_second_law}. At the phase transition, an energy cost $T S_\Gamma$ associated to the generation of degeneracies arises. Due to the dynamics depending only on the $z$-magnetization, the system presents no coherence.}
    \label{CurieWeiss}
\end{figure}

The Curie-Weiss Model describes an Ising magnet with infinite-range interactions through collective magnetization variables. It can be described (with respect to constants) by the Hamiltonian~\cite{CurieWeiss}
\begin{equation}
    H_t = -\frac{J}{N}J_z^2  - B_t J_z,
\end{equation}
where $J_z$ is the total magnetization along the $z$ direction, $J$ is a coupling constant (which we set to $1$ in our simulations), $N$ is the number of spins and $B_t$ is a time-dependent magnetic field. We then perform the following driving protocol: we initialize our system in a thermal Gibbs state at temperature $T = 0.5$ with $J = 1$, $N = 50$, and $B_0 = 2$ and reduce the magnetic field uniformly from $2$ at $t = 0$ to $0$ at $t = 5$. When the magnetic field reaches $B_5 = 0$, the system undergoes a phase transition where it's degeneracy structure changes, leading to an energy cost, as shown in Fig~\ref{CurieWeiss}.
%%%%%%%%%%%%%%%%%%%%%%%%%%%%%%%%%%%%%%%%%%%%%%%%%%%%%%%%%%%%%%%%%%%%%%%
%%%%%%%%%%%%%%%%%%%%%%%%%%%%%%%%%%%%%%%%%%%%%%%%%%%%%%%%%%%%%%%%%%%%%%%
It is very interesting to note that, since the degeneracy appears suddenly at $t = 5$, the bound in~\eqref{generalized_second_law} states that the work performed on the system up to that point must necessarily be enough to accommodate the energy cost $TS_\Gamma$.

%%%%%%%%%%%%%%%%%%%%%%%%%%%%%%%%%%%%%%%%%%%%%%%%%%%%%%%%%%%%%%%%%%%%%%%
%%%%%%%%%%%%%%%%%%%%%%%%%%%%%%%%%%%%%%%%%%%%%%%%%%%%%%%%%%%%%%%%%%%%%%%
%%%%%%%%%%%%%%%%%%%%%%%%%%%%%%%%%%%%%%%%%%%%%%%%%%%%%%%%%%%%%%%%%%%%%%%


\begin{thebibliography}{10}

\bibitem{Callen1991} H. B. Callen, {\em Thermodynamics and an Introduction to Thermostatistics} (Wiley, 1991).

\bibitem{Goold2016} J. Goold, M. Huber, A. Riera, L. del Rio, and P. Skrzypczyk, The role of quantum information in thermodynamics --- a topical review, J. Phys. A: Math. Theor. \textbf{49}, 143001 (2016).

\bibitem{Parrondo2015} J. M. R. Parrondo, J. M. Horowitz, and T. Sagawa, Thermodynamics of information, Nat. Phys. \textbf{11}, 131 (2015). 

\bibitem{Sagawa2013} T. Sagawa, \emph{Thermodynamics of Information Processing in Small Systems} (Springer, Tokyo, 2013).

\bibitem{Deffner2019} S. Deffner and S. Campbell, {\em Quantum Thermodynamics: An Introduction to the Thermodynamics of Quantum Information} (IOP Concise Physics, 2019).

\bibitem{Strasberg2022} P. Strasberg, {\em Quantum Stochastic Thermodynamics: Foundations and Selected Applications} (Oxford University Press, 2022).

\bibitem{Campbell2026} S. Campbell et al., Roadmap on quantum thermodynamics, Quantum Sci. Technol. \textbf{11}, 012501 (2026).

%\bibitem{Campbell2026} S. Campbell, I. D'Amico, M. A. Ciampini, J. Anders, N. Ares, S. Artini, A. Auff{\`e}ves, L. B. Oftelie, L. P. Bettmann, M. V. S. Bonan{\c c}a, T. Busch, M. Campisi, M. F. Cavalcante, L. A. Correa, E. Cuestas, C. B. Da{\u g}, S. Dago, S. Deffner, A. del Campo, A. Deutschmann-Olek, S. Donadi, E. Doucet, C. Elouard, K. Ensslin, P. Erker, N. Fabbri, F. Fedele, G. Fiusa, T. Fogarty, J. Folk, G. Guarnieri, A. S. Hegde, S. Hern{\'a}ndez-G{\'o}mez, C.-K. Hu, F. Iemini, B. Karimi, N. Kiesel, G. T. Landi, A. Lasek, S. Lemziakov, G. Lo Monaco, E. Lutz, D. Lvov, O. Maillet, M. Mehboudi, T. M. Mendon{\c c}a, H. J. D. Miller, A. K. Mitchell, M. T. Mitchison, V. Mukherjee, M. Paternostro, J. Pekola, M. Perarnau-Llobet, U. Poschinger, A. Rolandi, D. Rosa, R. S{\'a}nchez, A. C. Santos, R. S. Sarthour, E. Sela, A. Solfanelli, A. M. Souza, J. Splettstoesser, D. Tan, L. Tesser, T. Van Vu, A. Widera, N. Yunger Halpern, and K. Zawadzk, Roadmap on quantum thermodynamics, Quantum Sci. Technol. \textbf{11}, 012501 (2026).

\bibitem{ThermoGauge1} L. C. C\'eleri and \L . Rudnicki, Gauge-invariant quantum thermodynamics: Consequences for the first law, Entropy \textbf{26}, 111 (2024).

\bibitem{ThermoGauge2} G. F. Ferrari, \L . Rudnicki, and L. C. C\'eleri, Quantum thermodynamics as a gauge theory, Phys. Rev. A \textbf{111}, 052209 (2025).

\bibitem{Pernambuco2025} T. Pernambuco and L. C. C{\'e}leri, Geometric quantum thermodynamics: A fiber bundle approach, Phys. Rev. A \textbf{113}, 1103 (2026).

\bibitem{Esposito2009} M. Esposito, U. Harbola, and S. Mukamel, Nonequilibrium fluctuations, fluctuation theorems, and counting statistics in quantum systems, Rev. Mod. Phys. \textbf{81}, 1665 (2009).

\bibitem{Jarzynski2011} C. Jarzynski, Equalities and inequalities: Irreversibility and the second law of thermodynamics at the nanoscale, Annu. Rev. Condens. Matter Phys. \textbf{2}, 329 (2011).

\bibitem{Seifert2012} U. Seifert, Stochastic thermodynamics, fluctuation theorems and molecular machines, Rep. Prog. Phys. \textbf{75}, 126001 (2012).

\bibitem{Campisi2011} M. Campisi, P. Hänggi, and P. Talkner, Colloquium: Quantum fluctuation relations: Foundations and applications, Rev. Mod. Phys. \textbf{83}, 771 (2011)

\bibitem{Jarzynski1997} C. Jarzynski, Nonequilibrium equality for free energy differences, Phys. Rev. Lett. \textbf{78}, 2690 (1997).

\bibitem{Jarzynski1997b} C. Jarzynski, Equilibrium free-energy differences from nonequilibrium measurements: A master-equation approach, Phys. Rev. E \textbf{56}, 5018 (1997)

\bibitem{Crooks1998} E. Crooks, Nonequilibrium measurements of free energy differences for microscopically reversible Markovian systems, J. Stat. Phys. \textbf{90}, 1481 (1998).

\bibitem{Crooks1999} G. E. Crooks, Entropy production fluctuation theorem and the nonequilibrium work relation for free energy differences, Phys. Rev. E \textbf{60}, 2721 (1999).

\bibitem{Crooks2000} G. E. Crooks, Path-ensemble averages in systems driven far from equilibrium, Phys. Rev. E \textbf{61}, 2361 (2000)

\bibitem{Note1} In this Letter, we choose to work with energy measurements, but the theory works for any other observable (or set of observables); The structure of the symmetry group will change, as well as the physical meaning of the involved quantities; However, the mathematical form of the theory and the geometric interpretations of equations will remain the same.

\bibitem{Note2} In a general situation, we must employ the time-reverse operator to define the reverse evolution, but in most situations of interest, it suffices to consider that the reverse evolution is governed by the unitary operator $U_\tau^\dagger$. We consider this case here, although the extension to more general cases is straightforward.

\bibitem{Alicki1979} R. Alicki, The quantum open system as a model of the heat engine, J. Phys. A: Math. Gen. \textbf{12}, L103 (1979).

\bibitem{Micadei2013} K. Micadei, R. M. Serra, and L. C. Céleri, Thermodynamic cost of acquiring information, Phys. Rev. E \textbf{88}, 062123 (2013).

\bibitem{Bouton2021} Q. Bouton, J. Nettersheim, S. Burgardt, D. Adam, E. Lutz and A. Widera, A quantum heat engine driven by atomic collisions, Nat Commun \textbf{12}, 2063 (2021)

\bibitem{Pijn2022} D. Pijn, O. Onishchenko, J. Hilder, U. G. Poschinger, F. Schmidt-Kaler, and R. Uzdin, Detecting heat leaks with trapped ion qubits, Phys. Rev. Lett. \textbf{128}, 110601 (2022).

\bibitem{Yan2018} L. L. Yan, T. P. Xiong, K. Rehan, F. Zhou, D. F. Liang, L. Chen, J. Q. Zhang, W. L. Yang, Z. H. Ma, and M. Feng, Single-atom demonstration of the quantum Landauer principle, Phys. Rev. Lett. \textbf{120}, 210601 (2018).

\bibitem{Onishchenko2024} O. Onishchenko, G. Guarnieri, P. Rosillo-Rodes, D. Pijn, J. Hilder, U. G. Poschinger, M. Perarnau-Llobet, J. Eisert and F. Schmidt-Kaler, Probing coherent quantum thermodynamics using a trapped ion, Nat Commun \textbf{15}, 6974 (2024).

\bibitem{Oliveira2020} A. G. de Oliveira, R. M. Gomes, V. C. C. Brasil, N. Rubiano da Silva, L. C. Céleri, and P. H. Souto Ribeiro, Full thermalization of a photonic qubit, Phys. Lett. A \textbf{384}, 126933 (2020).

\bibitem{Ribeiro2020} P. H. Souto Ribeiro, T. Häffner, G. L. Zanin, N. Rubiano da Silva, R. Medeiros de Araújo, W. C. Soares, R. J. de Assis, L. C. Céleri, and A. Forbes, Experimental study of the generalized Jarzynski fluctuation relation using entangled photons, Phys. Rev. A \textbf{101}, 052113 (2020).

\bibitem{Zanin2019} G. L. Zanin, T. Häffner, M. A. A. Talarico, E. I. Duzzioni, P. H. Souto Ribeiro, G. T. Landi, and L. C. Céleri, Experimental quantum thermodynamics with linear optics, Braz. J. Phys. \textbf{49}, 783 (2019).

\bibitem{Chen2019} Y.-Y. Chen, G. Watanabe, Y.-C. Yu, X.-W. Guan, and A. del Campo, An interaction-driven many-particle quantum heat engine and its universal behavior, npj Quantum Inf \textbf{5}, 88 (2019).

\bibitem{Koch2023} J. Koch, K. Menon, E. Cuestas, S. Barbosa, E. Lutz, T. Fogarty, T. Busch, and A. Widera, A quantum engine in the BEC–BCS crossover, Nature \textbf{621}, 723 (2023).

\bibitem{Watanabe2020} G. Watanabe, B. P. Venkatesh, P. Talkner, M.-J. Hwang, and A. del Campo, Quantum statistical enhancement of the collective performance of multiple bosonic engines, Phys. Rev. Lett. \textbf{124}, 210603 (2020).

\bibitem{Costa2026} R. L. S. Costa, M. L. W. Basso, J. Maziero, and L. C. Céleri, Work distribution of quantum fields in static curved spacetimes, Phys. Rev. D \textbf{113}, 025010 (2026).

\bibitem{Moreira2024} T. H. Moreira, L. C. Céleri, Entropy production due to spacetime fluctuations, Class. Quantum Grav. \textbf{42} 025022 (2024).

\bibitem{Basso2025} M. L. W. Basso, J. Maziero, and L. C. Céleri, Quantum detailed fluctuation theorem in curved spacetimes: The agent dependent nature of entropy production, Phys. Rev. Lett. \textbf{134}, 050406 (2025).

\bibitem{Basso2023} M. L. W. Basso, J. Maziero, and L. C. Céleri, The irreversibility of relativistic time-dilation, Class. Quantum Grav. \textbf{40}, 195001 (2023).

\bibitem{Spohn} H. Spohn, Entropy production for quantum dynamical semigroups, J. Math. Phys. \textbf{19}, 1227 (1978).

\bibitem{Baumgratz2014} T. Baumgratz, M. Cramer, and M. B. Plenio, Quantifying coherence, Phys. Rev. Lett. \textbf{113}, 140401 (2014).

\bibitem{Bartlett2003} S. D. Bartlett and H. M. Wiseman, Entanglement constrained by superselection rules, Phys. Rev. Lett. \textbf{91}, 097903 (2003).

\bibitem{Marvian2014} I. Marvian and R. W. Spekkens, Extending Noether’s theorem by quantifying the asymmetry of quantum states, Nat. Commun. \textbf{5}, 3821 (2014).

\bibitem{DeffnerLutz} S. Deffner and E. Lutz, Thermodynamic length for far from equilibrium quantum systems, Phys. Rev. E. \textbf{87}, 022143 (2013).

\bibitem{GeometryOfQuantumStates} I. Bengtsson and K. \.{Z}yczkowski, \emph{Geometry of Quantum States: An Introduction to Quantum Entanglement} (Cambridge University Press, Cambridge, UK, 2017).

\bibitem{LandauZener} C. Zener, Non-adiabatic  crossing of energy levels, Proc. R. Soc. Lond. A \textbf{137}, 833 (1932).

\bibitem{PartialLandauZener} J. R. F. Lima and G. Burkard, Partial Landau-Zener transitions and applications to qubit shuttling, Phys. Rev. B \textbf{111}, 235439 (2025).

\bibitem{CurieWeiss} M. Kochmański, T. Paszkiewicz, and S. Wolski, Curie–Weiss magnet---a simple model of phase transition, Eur. J. Phys. \textbf{34}, 155 (2013).

\end{thebibliography}
\end{document}